# Scientific Programs Imply Uncertainty.
# Results Expected and Unexpected.

> Science and engineering have requests for a wide variety of programs, but I think that all of them can be divided between two groups.  Programs of the first group deal with the well known situations and, by using well known equations, give results for any combination of input parameters.  Such programs are specialized very powerful calculators.  Another group of programs is needed to analyse the situations with different levels of uncertainty.  Programs are developed at the best level of their authors, but scientists need to look at the situations beyond the area of current knowledge, and they need programs to do analysis in the areas of uncertainty.  Is it possible do design programs which allow to analyse the situations beyond the knowledge of developers?

50 years ago in his famous book *The Structure of Scientific Revolutions* [1] Thomas Kuhn described two different ways in which science goes on and on.  At each moment every branch of science has its laws which are considered to be correct without any doubts; studies in universities and work in laboratories go according to these laws.  Knowledge is gained by applying these laws to some new experiments or by checking again and again some previously described experiments in order to prove their correctness or fault.  Years of meticulous work can cast doubt on the existing theory; then some new experiments and the analysis of new results under different angle can bring out the new revolutionary theory which may replace the previous one and become the new law.

Kuhn wrote his book long before the era of personal computers at the time when big computers were used as very powerful calculators in several (limited enough) areas.  Later personal computers were introduced; eventually they became the mandatory part of equipment.  It is not only a piece of equipment as any other device in physical experiment.  A lot of data and the whole process of analysis depend on the way the data is obtained, processed, and demonstrated (visualized), so the results comprehension depends on the way the used programs are developed.

Several years ago I published in CiSE an article about the way in which movability of all the screen objects changes the design of programs and the way users (especially scientists and engineers) can work with the programs of the new type [2]; much more on these topics can be found in [3, 4].  The mentioned publications use for illustration different programs which are focused mostly on the analysis of previously obtained results and visualization, so I would call such programs calculators with unlimited possibilities of visualization.  There are other applications which are highly requested in science and engineering.  These are the programs with high level of uncertainty.  At the stage of development, it is often impossible to formulate all the requirements for such programs, so they must be designed in such way that it would be possible for users to analyse the situations about which developers were not even thinking.

Some scientific experiments, especially in physics, can be very expensive, may need a lot of time and efforts for preparation, and can be even dangerous.  Some of experiments cannot be organized on Earth.  Wide spread of computers and the achievements in programming brought to the surface the idea of switching from real experiments to their modeling.

In organizing real experiments, you are free in your decisions about using the equipment and all the involved elements.  While working with the modeling program designed by somebody else, you are absolutely limited by its designer's view on the proposed model and you can't go out of this view.  You can do only whatever the developer of this model considered to be correct.  As a rule, scientists in each particular area are much better specialists in their area than developers of the programs they have to use.  This difference in levels of knowledge and understanding causes the paradoxical situation when scientists, in their search for new and unexpected, are limited by the vision of lesser specialists.  User-driven applications help to find the way out of this programming dead-end.

Whenever I come out with any suggestion, I prefer to demonstrate it with some working programs.  The explanation becomes much more obvious if a demonstrated program deals with some item or area which nearly everyone is familiar with.  I hope that nearly every reader of this article is familiar with the main postulates and elements of Geometric Optics.  A small book on Geometric Optics [5] is accompanied by a program which makes the understanding of the discussed questions much easier.  Further on the article is illustrated with the examples and figures from this book, but I want to underline that conclusions are applicable to all other areas of science and engineering.

## From expected results to unexpected

One doesn't need to go half a world around to see something amazing.  It is often enough to make a couple of steps from the familiar path and to look at the surrounding scenery at a new angle.  The same happens in experiments and programs (!) if such small steps aside are allowed by design and you wish to make them.  Unexpected results may occur even in a trivial situation and you always feel amazed when such thing happens.

Refraction of light is described by the Snell's Law and is often illustrated with a ray going through glass plate or through air – water boundary. These situations are often observed in our every day life, so we get the explanations of the familiar things. While looking through a window not directly but at some angle, we have a chance to observe that the outside picture behind the window is not distorted but only shifted a bit. There is also a canonical case of refraction when we look from the river (or lake) bank on the water and see the distorted view of anyone staying in the water. You can find these two cases as illustrations in nearly all textbooks on Geometric Optics.

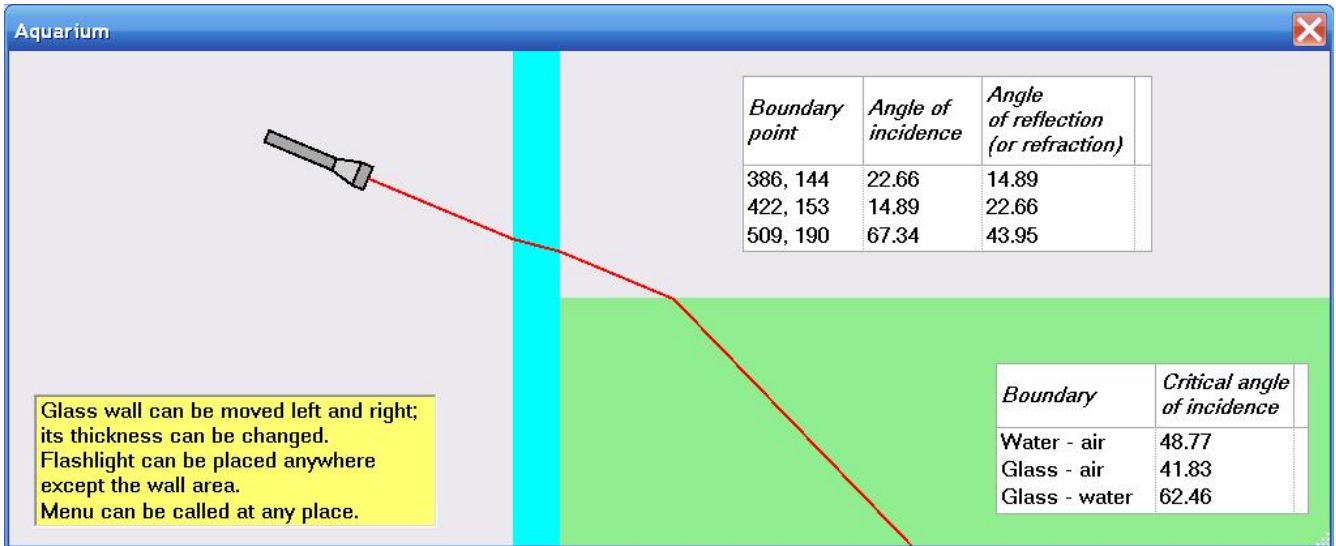

**Fig.1** Glass wall goes from bottom to top while water fills only half of the volume on one side.

In order to make the discussion more interesting, I included an example of a big aquarium (oceanarium). This single example allows to analyse air – glass, air – water, and glass – water boundaries; each boundary can be crossed in both directions, so the total number of variants is six. In a classical fixed design of a program, it would be enough to discuss the process of refraction with the situation from **figure 1**. In the current example, with all the elements movable, you can easily move and rotate the flashlight, and by doing this you can obtain absolutely unexpected situations which will tell you much more about refraction. I am sure that some results will amaze you, but the Snell's Law is universal, and up till now it was not refuted.

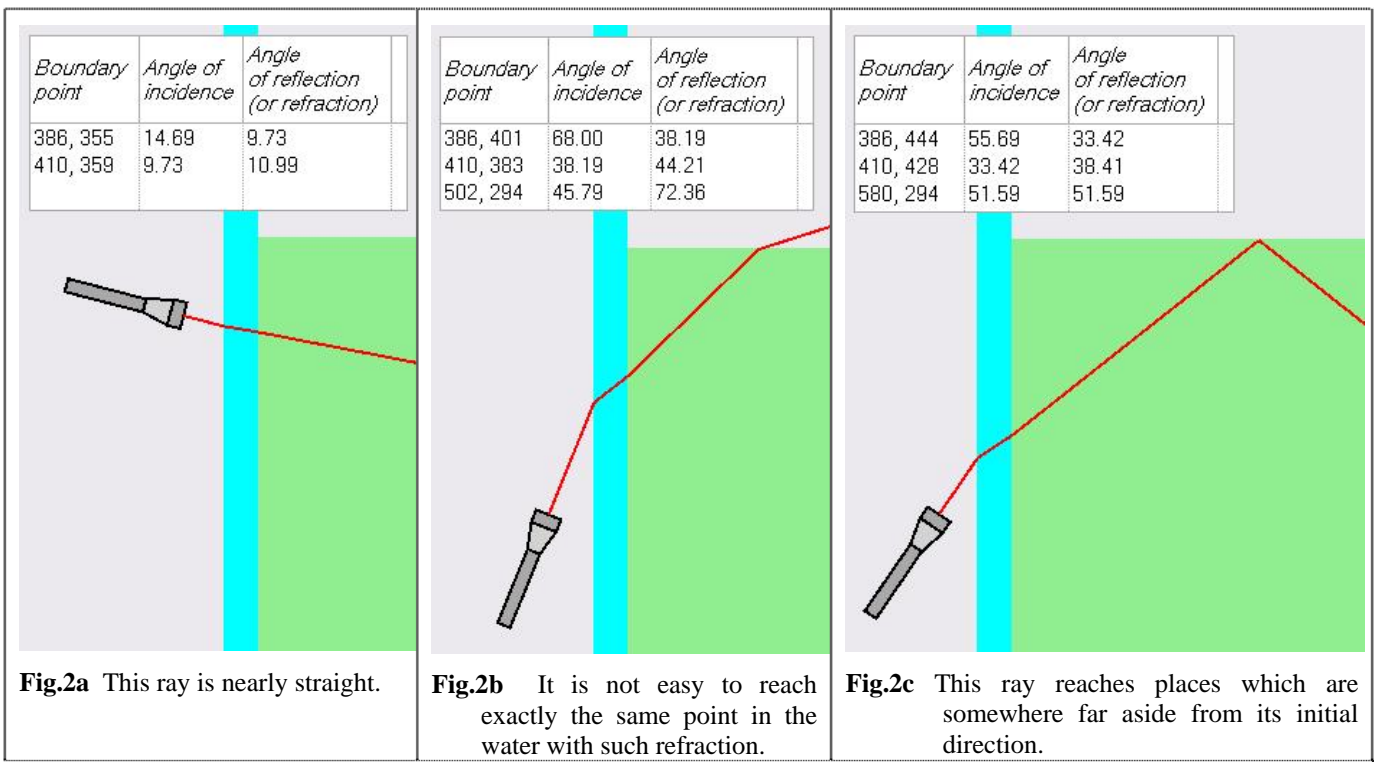

**Fig.2a** This ray is nearly straight.　　**Fig.2b** It is not easy to reach exactly the same point in the water with such refraction.　　**Fig.2c** This ray reaches places which are somewhere far aside from its initial direction.

Suppose that parents with small kids are on a visit to oceanarium; difference in height is important for further discussion. Family members stay near the glass wall and look for interesting creatures on the other side. A grown up sees something amazing in front (**figure 2a**) and tells kids to look there. They try, but in order to see the mentioned creature they have to look at absolutely different angle (**figure 2b**). Maybe they will see the same creature, but chances are high that they miss because of those turns of the ray at both sides of the glass wall. Parent insists that there is really something amazing just in front, so kids, in an attempt to find it, slightly change the angle of their look. Instead of something just near the glass wall they have a chance to see creatures in the far away corner at the bottom of oceanarium (**figure 2c** doesn't show the full size of oceanarium because of the page limit, but you can easily prolong the last leg of the ray route). Kids were not even thinking about that far away part of oceanarium but they can see it.

It is not an artificial situation which I thought out in order to make the text more interesting. **Figure 2c** reminded me of one unsolved problem which puzzled me many years ago. In my childhood, we had an aquarium placed on top of a bookcase. The bookcase was nearly twice higher than I was at that time. In order to see the fish, I had to look up and then I could see the fish just behind the glass. If I stood some distance from the bookcase and looked up, then occasionally I could see the underwater cave in the farthest corner of aquarium. I couldn't see it directly because its view was blocked by the upper part of the bookcase. This occasional view of the hidden corner puzzled me, but it was years before I learnt about the total internal reflection. Now years later I can easily explain that episodes from the childhood. Those strange pictures of the hidden aquarium corners were stored somewhere deep inside my memory for many years and came back only when the program showed me route from **figures 2c**.

With the help of this program I am trying to demonstrate that if users get total control over all the screen elements, a very small change initiated by user allows to move from expected results to unexpected. In science, both types of results can be very valuable. Expected results allow to prove the correctness of suppositions made before the experiment. Unexpected results show that the whole theory on which experiments are based can be partly or entirely wrong. The same experiment (**figure 1**) allowed me to see other results about which I couldn't make any predictions before I designed this example.

Can you say what the underwater creatures see when they look from inside the oceanarium? In order to avoid the discussion of nonhuman mind, let us assume that there are mermaids performing inside or scientists / maintenance workers doing something. Can they see visitors behind the glass wall? I don't think that too many visitors burden themselves with such questions. In our everyday life, if we see somebody, then this *somebody* can see us. The reversibility of light means that if it is sent in the opposite direction it goes back by the same route. If we stay near the glass wall and watch different creatures on the other side then those creatures can see us. The law always works if a creature looks exactly along the ray route in the opposite direction, but what happens if the angle is different?

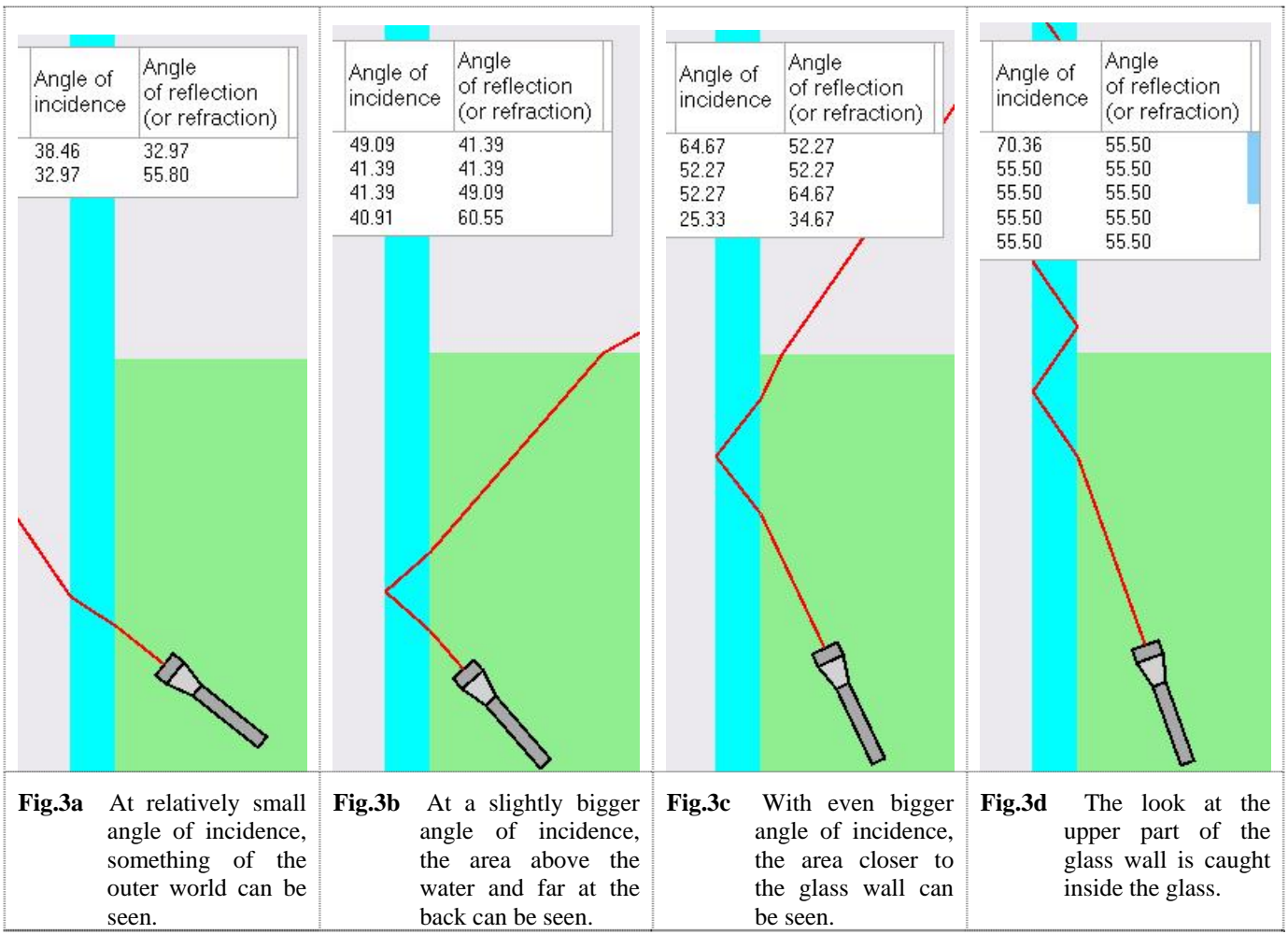

**Fig.3a** At relatively small angle of incidence, something of the outer world can be seen.

**Fig.3b** At a slightly bigger angle of incidence, the area above the water and far at the back can be seen.

**Fig.3c** With even bigger angle of incidence, the area closer to the glass wall can be seen.

**Fig.3d** The look at the upper part of the glass wall is caught inside the glass.

**Figures 3** show route variants when somebody looks from under the water at different angles to the glass wall. I would say that only result from **figure 3a** is expected while three others are unexpected. When creature looks at relatively small angle of incidence (**figure 3a**), there is some distortion caused by refraction on two boundaries, but at least something on the outside can be seen. With the bigger angle of incidence, the underwater creature has no chances to see visitors. **Figures 3b, 3c,** and **3d** show variants for the steadily increasing angle of incidence; I wonder how many readers would expect such results. I wasn't thinking about such results and you would never see them at all if not for movability and rotatability of a flashlight in the current example.

From time to time scientists receive in their experiments some unexpected results. Often enough these strange results are considered as the equipment malfunction and are thrown away. In rare cases such strange results led to thorough research which ended with Nobel Prize award and to famous inventions. (X-rays discovery in 1895 and microwave oven invention in 1945, to mention the few).

Science is an exploration of the unknown things. Users of scientific applications developed in a standard way (all currently used programs) can see only the results which are approved by designers; thus, unexpected results are simply impossible. If a scientific program substitutes real experiments, then it must allow to see unexpected results. If such thing is forbidden by the main design idea, then such program is called a scientific only by mistake. It is really a very serious question: "*Is it possible at all to design a scientific program which allows to receive the unexpected results?*"

Let us consider one more example (or better to say set of examples) which demonstrates a result… You will decide yourself, whether it is expected or unexpected.

350 years ago Isaac Newton demonstrated his famous experiments with triangular prism in order to explain the origin of light. Since then the picture of a triangular prism with white ray touching one plane and a set of seven colored rays (or cones) emerging from another plane can be found in every book on Geometric Optics. To demonstrate that the difference between expected and unexpected is infinitesimal, I can ask a very simple question: "Is it possible to see less than seven colored cones"? (Book [5] with an example of triangular prism allows to find an answer.)

Seven colored cones emerging from triangular prism is a perfect demonstration of light dispersion. Another, even more famous example of such dispersion is rainbow. To explain rainbow phenomenon, the ray propagation through a water droplet is used. Water droplet is circular. Circle can be considered as a regular polygon with the infinitive number of vertices (or planes). Thus we observe light dispersion in triangular prism and we observe dispersion (rainbow) with the help of prisms which have an infinitive number of planes. (To be correct, rainbow is the result of dispersion on a set of droplets with each colored ray coming from different droplet.) What happens when the white light comes to a regular polygon with a number of planes more than three? What happens when this number is changed?

I like to investigate new areas. The transformation of the screen triangular prism into another one with the changeable number of planes was done quickly. I could see the results similar to triangular prism, so at first such prism transformation didn't produce any unexpected results. Then I began to watch something absolutely different; after some time I understood that these unexpected results were described in a famous *Pollyanna* book by Eleanor H. Porter [6].

"*Pollyanna had not hung up three of the pendants in the sunlit window before she saw a little of what was going to happen. She was so excited then she could scarcely control her shaking fingers enough to hang up the rest. But at last her task was finished, and she stepped back with a low cry of delight.*

*It had become a fairyland--that sumptuous, but dreary bedroom. Everywhere were bits of dancing red and green, violet and orange, gold and blue. The wall, the floor, and the furniture, even to the bed itself, were aflame with shimmering bits of color.*"

In case of equilateral triangular prism, the maximum angle between red and violet parts of the dispersed light is somewhere around 10 degrees, so, the light coming out from such prism would place the colored spots next to each other. The vivid description from the book tells about the colored spots all round the room, so the cause of such picture must be different; **figures 4** show how it happens. To avoid the mess of all different colors on one figure, I prepared three different figures for different light components. The geometry of the involved elements is the same for all three. Now imagine that sun light touches the pendant in the way it is shown on the figures. The white light is dispersed; then each component will have its own way and will come out at its own angle; as a result we get different colored spots in absolutely different places.

*Pollyanna* book became famous more than a century ago; and since then it was read and admired by millions of people. In many houses the polygonal pendants are hanged in the way of a sun light, so people observe similar fairy pictures. Among those millions of people, there were (and are) thousands and thousands of specialists in physics and optics. Why did it happen so that I never saw or heard any explanation of this phenomenon? Did it happen only because each one of those specialists decided it not worth mentioning? I wonder.

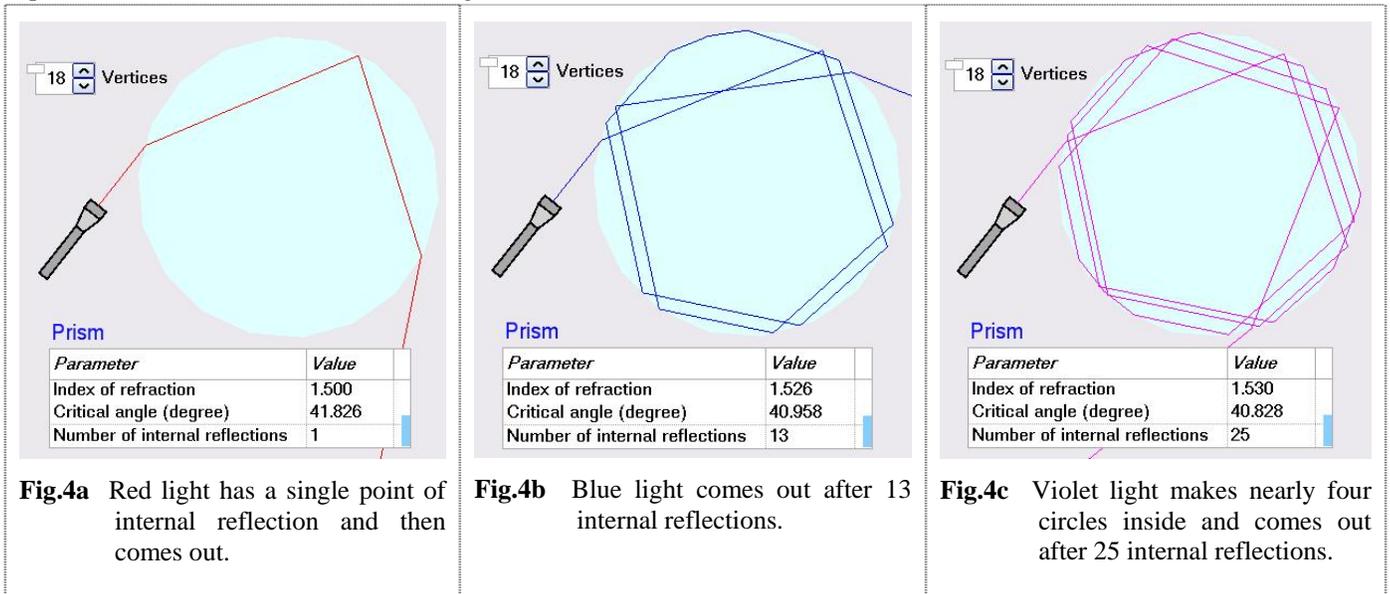

**Fig.4a**  Red light has a single point of internal reflection and then comes out.

**Fig.4b**  Blue light comes out after 13 internal reflections.

**Fig.4c**  Violet light makes nearly four circles inside and comes out after 25 internal reflections.

Did I expect to see such result when I designed an example with regular polygonal prism? Certainly not, though I was aware of gem cutting used for centuries. (We may skip the discussion of my ability to predict the results, but at least I am honest.) Luckily, the program was designed as a user-driven application, so user (me!) could do whatever he wanted with the involved elements. As usual, this user decided to play with the screen elements and on the way received the results which designer (also me!) would never think about. It is a perfect answer to the fundamental question: "*Is it possible to receive the unexpected results from the scientific program?*"

## Conclusion

I developed a user-driven application for the area of Geometric Optics and received very interesting results. I am sure that, by playing with this program, users can obtain even more interesting results than I have demonstrated. Designer of such application cannot declare the full list of possible results; everything depends on how users work with an application and on users' creative ability.

If you design user-driven applications for other areas of science and engineering you'll receive not less interesting results for those areas.

Currently developed programs for science and engineering always put a lot of restrictions on their use. As a rule, these limitations were taught to application designers 5 – 10 – 15 earlier as the prevailing ideas at that time; so the programs are years behind the things which scientists try to explore at the current moment.

The main goal of designing applications for science and engineering is not receiving money; it is a good result but not the goal. The main goal is the design of instruments with which scientists and engineers can explore and analyse the problems and new effects. I don't care how long it will take to get away from the restrictions of adaptive interface and to switch to the ideas of user-driven applications. It would be the only way to design programs which allow users to explore new things and to receive the unexpected results which in science are usually the most important.

When Newton and Leibniz independently invented calculus, they didn't tour Europe in order to persuade every scientist to use a new instrument. On the contrary, each scientist had to decide for himself whether it worth spending time on reading about calculus and understanding it. Then scientists had to think out the way to apply this new and very powerful math instrument to their particular research work. It happened roughly 300 years ago when the number of scientists was comparatively small, each one, as a rule, worked not in some very limited area but on a number of problems which were far away from each other, and nearly every scientist had a solid math education. As a result, the use of calculus spread to nearly all areas where math was used. (Can you name any area where it is not used?) Since then there were no discussions whether to use calculus or not.

I am not going to demonstrate what user-driven applications can do in each particular area. It is for programmers in all areas to understand the ideas of user-driven applications and, if they want to try, to apply them to their work. Examples make this process easier, but all depends on one's wish and ability to use new ideas.